\documentclass[letterpaper,11pt]{article}
\pdfoutput=1

\usepackage{jheppub}
\usepackage{booktabs} 
\usepackage{graphicx} 
\usepackage{siunitx}
\usepackage{cleveref}
\usepackage{epsfig}
\usepackage[normalem]{ulem}
\usepackage{bm}
\usepackage{bbm}
\usepackage{slashed}
\usepackage{xspace}
\usepackage{subfigure}
\usepackage{multirow}

\newcommand{\setmainskip}{\setlength\baselineskip{18pt}}

\usepackage{amsmath,latexsym}
\usepackage{pstricks}
\usepackage{color} 
 
\usepackage{marginnote}

\newcommand{\nn}{\nonumber}

\newcommand{\beq}{\begin{equation}}
\newcommand{\eeq}{\end{equation}}
\newcommand{\bea}{\begin{eqnarray}}
\newcommand{\eea}{\end{eqnarray}}

\allowdisplaybreaks[1]

\begin{document}
							
							
							
							
\title{\boldmath  Symmetry Realization via a Dynamical Inverse Higgs Mechanism }

\author{Ira Z. Rothstein$^1$ and  Prashant Shrivastava $^1$}
							
\affiliation{$^1$Department of Physics, Carnegie Mellon University, Pittsburgh, PA 15213, USA}
\emailAdd{izr@andrew.cmu.edu}
\emailAdd{prashans@andrew.cmu.edu}
							
\abstract{
								
\vspace{0.2cm}
\setlength\baselineskip{15pt}
							
The  Ward identities associated with spontaneously broken symmetries can be saturated
by Goldstone bosons. However, when space-time symmetries
are broken, the number of Goldstone bosons necessary to non-linearly realize the symmetry
can be less than the number of broken generators. 
The loss of Goldstones may be due to a  redundancy or the generation of  a gap. In either case the associated Goldstone may be removed from the spectrum.
This phenomena is called   an Inverse Higgs Mechanism (IHM) and its appearance has a well defined mathematical condition. However, there are cases when a Goldstone boson associated with a broken
generator  does not appear in the low energy theory despite the lack of the existence of an associated IHM. In this paper we will show that in such cases the relevant broken symmetry can be realized, without the aid of an associated Goldstone,  if there exists a proper set of
 operator constraints, which we call a Dynamical Inverse Higgs Mechanism (DIHM).  
 We consider the spontaneous breaking of  boosts, rotations and conformal transformations in the context of Fermi liquids, finding three possible
 paths to symmetry realization:  pure Goldstones, no Goldstones and  DIHM, or some mixture thereof.  We show that in the two dimensional
 degenerate electron system  the DIHM route  is the only consistent way to realize spontaneously broken boosts and dilatations,  while in three dimensions these symmetries could just as well be realized via the inclusion of non-derivatively coupled
 Goldstone bosons. We present the action, including the leading order non-linearities, for the rotational Goldstone (angulon), and discuss
the constraint associated with the possible DIHM that would need to be imposed to remove it from the spectrum.
Finally we discuss the conditions under which  Goldstone bosons are non-derivatively coupled, a necessary condition for the existence of a Dynamical Inverse Higgs Constraint (DIHC), generalizing the results for Vishwanath and Wantanabe.

}
							
							
							
\maketitle
\setmainskip

							
							

\newpage

\section{Introduction}

 When  symmetries
are broken spontaneously  they are manifested non-linearly in the IR. 
The realization of the broken symmetry, in general, will include gapless Goldstone bosons (GBs).
When space-time symmetries are unbroken, Goldstone bosons are derivatively coupled and are irrelevant in the IR. If there are other gapless modes in the spectrum, not associated with symmetry breaking, the
Goldstones may be ignored to first approximation.
The canonical example of such a scenario is the  Fermi 
liquid theory of metals  where phonons do not play a role at leading order\footnote{Potential, off-shell phonons play an indirect
role in that they contribute to the attractive piece of  the four Fermi coupling once they have been integrated out. }.
Of course, if there are no other gapless modes, or if the Goldstones couple to sources, then  they are
of primary importance. An example of such a scenario is the QCD chiral Lagrangian.

 When space-time symmetries are broken, GBs
 can be non-derivatively coupled. Two canonical examples being the relativistic dilaton and the Goldstone
bosons of broken rotational invariance in Fermi liquids. Such non-derivative couplings   lead to marginal or relevant  interactions which can drastically
affect the IR physics. For instance, when rotational invariance is broken
in a Fermi liquid and translations are unbroken (nematic order),  the quasi-particles decay into Goldstones   \cite{OKF,LBFFO,X} leading to a  width which scale as $\Gamma \sim E^{\alpha}$ with $\alpha <2$.
While relativistic dilatons can generate  long range forces  and, as such, their couplings are highly constrained\cite{Kaplan:2000hh}. The necessary conditions for non-derivatively coupled Goldstones in non-relativistic  theories were discussed by Vishwanath and Watanabe in \cite{VW}.


For relativistic theories the breaking of internal symmetries leads to a one to one correspondence
between Goldstone bosons and generators which shift the vacuum \cite{Goldstone:1961eq,Goldstone:1962es}. Moreover, the Goldstone boson
is manifested as a delta function in the spectral density. 
When space-time symmetries are broken\cite{Brauner:2010wm} (this includes the aforementioned non-relativistic case)  we are no longer assured about the existence of a Goldstone boson
associated with  
 a broken generator $X$.   Suppose we have a order parameter $\phi$ such that 
\beq
\langle \Omega \mid [X,\phi] \mid \Omega \rangle \neq 0,
\eeq
where \beq
X= \int d^{d-1}x j_0^X(x).
\eeq
It follows that  \footnote{We assume here that there are no long range forces so that surface terms may be dropped and that generators have canonical translational properties. See \cite{Brauner:2010wm} for a discussion.}
\beq
\label{res}
 \sum_n \delta^{(d-1)}(\vec p_n) \left[ \langle \Omega \mid j^X_0(0) \mid n \rangle \langle n \mid \phi(0)\mid \Omega \rangle e^{iE_nt}- \langle \Omega \mid \phi(0)\mid n \rangle \langle n \mid j^X_0(0) \mid  \Omega \rangle e^{-iE_nt}\right]\neq 0.
\eeq
We assume that the system preserves a discrete translational invariance, so that there exists some notion of a conserved momentum.
Given that $X$ is a conserved charge, we see that symmetry breaking implies the existence of a zero energy state
when $\vec p_n \rightarrow 0$. 
However, we can not say anything about the associated spectral weight other than the fact that it has to non-zero.
This state may be arbitrarily wide.
Thus if we are to count Goldstone bosons when space-time symmetries are broken we must
define what we mean by a Goldstone boson.
For our purposes we will define a Goldstone mode as having to satisfy the definition of a quasi-particles,
$\Gamma \leq E^2$ in the limit of vanishing energy. Also note that (\ref{res}) does not preclude the
possibility of having multiple gapless states.

For non-relativistic systems, there can be no symmetry breaking if the vacuum is trivial since pair
creation is disallowed.
Thus a non-relativistic system which manifests {\it any} symmetry breaking necessarily has a ground
state which breaks at least boost invariance and one can not
separate space-time from internal symmetry breaking. However, in the literature when internal
symmetries are broken, the breaking of boost symmetry is usually ignored. We will come back to this important  issue below. 


Goldstone bosons may have various dispersion relations. Inequalities for 
counting rules for the type I ($E \sim p$) and type II ($E\sim p^2$)  Goldstones
\footnote{The dispersion relation need not be limited to these two choices. Higher order relations are possible in some systems , e.g. the vibrations of a stiff rod. Non-analytic dispersion relations such as those discussed in
\cite{WM} are due to integrating out fields with analytic dispersion relations.} 
 associated with internal symmetry breaking were first written down by
 Nielsen and Chadha \cite{Nielsen:1975hm}. Since then a series of papers ultimately led to the final result for
 the number ($N$)  of Goldstones \cite{Leutwyler:1993gf,Nambu:2004yia,Schafer:2001bq,Watanabe:2011ec,Watanabe:2012hr,Hidaka:2012ym} when the group ${\rm G}$ is broken to ${\rm H}$\cite{WM}
\beq
\label{result}
N= {\rm dim(G/H)-\frac{1}{2}} {\rm rank} [\rho]
\eeq
where $\rho_{ab}= -i \langle [X_a,j^a_b(0)]\rangle$ and $X_a$ are the broken charges of the full group $\rm G$ and $j^b(0)$ are the associated  charge densities. Furthermore counting rules for gapped Goldstones (with both calculable
and incalculable gaps ) have been developed \cite{gapped,anton}. 

The analysis leading to the result (\ref{result}) does not hold when  space-time symmetries are broken.
Consider the case of a canonical superfluid. This system breaks a $U(1)$ symmetry corresponding to particle
number and  the rank of $\rho$ vanishes leading to a prediction of one Goldstone boson.
However, we must ask what justifies ignoring the ersatz GB arising from the breaking of boost invariance?
The answer lies in whats known as the 
as the ``Inverse Higgs Mechanism''\cite{Ivanov}(IHM)  (see also \cite{ML}). 

The counting of (gapless) Goldstones  still follows
once we have
established the necessary criteria for the IHM .   When two broken generators $X,X^\prime$ obey a relation of the form
\beq
\label{IH}
[\bar P,X] \propto X^\prime 
\eeq
where $\bar P$ are the unbroken translations and $X$ and $X^\prime$ are not in the same
$ \rm H$ multiplet, it {\it may} be possible to  eliminate the Goldstone associated with $X$. 
As emphasized in \cite{gapped} the algebraic relation (\ref{IH}) may or may not be the  signal of a redundancy. 
That depends upon the nature of the order parameter. In particular, given a set of  broken generators $X_a$  a redundancy exists when there
is a non-trivial solution to the equation
\beq
\pi^a(x) X^a \langle O(x) \rangle=0
\eeq
where $\langle O(x) \rangle$ is the order parameter. As an example consider the symmetry breaking pattern for
a metal. The lattice breaks rotations, translations and boosts. The boost Goldstone $\eta^i$ and rotation Goldstone $\theta^{i}$ can be easily seen to be
redundant since 
\beq
K^i = P^i t -M x^i
\eeq
and
\beq
J^i = \epsilon^{ijk} x^jP^k
\eeq
thus, assuming that the mass $M$ is unbroken, i.e. no condensation, we have
\beq
(\theta^k(x) \epsilon^{ijk} x^i (i \partial^j)+\eta^i(x) (it\partial^i ) + \pi^a(x) (i\partial^a)) \langle O(x) \rangle=0
\eeq
so that both rotations and boosts can be compensated for by a Goldstone dependent translation \cite{ML}.

In any case when  condition (\ref{IH}) is satisfied, it is often possible to impose a constraint
on the fields which is consistent with the symmetries. This constraint is called the Inverse Higgs Constraints (IHC)
which is associated with the IHM.

\subsection{The Missing Goldstones}
As was pointed out in \cite{framids} there are cases for which there is no inverse Higgs constraints
and yet the Goldstones still do not appear.  If particle number is spontaneously broken, then due to the fact that
$[P,K] \propto M$, there is an IHC which allows one to eliminate the boost Goldstone.
But if there is no IHM involving the boost generator, one must include the
boost Goldstone in the analysis.  In \cite{framids} the authors considered  two such symmetry breaking
patterns called type-I  and type-II ``framids''. The former is a system  in which  the only broken symmetry is boost invariance while
the latter also breaks rotations. A cursory check of the Galilean algebra shows that none of the broken generators
satisfy (\ref{IH}) in these cases and yet  the  Goldstones associated with boosts, dubbed the ``framons", are nowhere to be seen
in nature. 

Another missing Goldstone boson arises in the case of non-relativistic dilatation invariance. The authors of  \cite{oz} point out that given that
the dilaton $ \sigma$ transforms under Galilean boosts as
\beq
\sigma(x,t) \rightarrow \sigma(x-vt,t)
\eeq
there is no way to write down a boost invariant kinetic term\footnote{Matter fields transform as projective representations under boosts which allow for the canonical kinetic energy term.} since the
time derivative of the dilaton transforms non-trivially.
As also pointed out in \cite{oz}, if the $U(1)$ of particle number is broken then, as a consequence of
the algebraic relation,
\beq
\label{U1}
[P_i,K_j]=i\delta_{ij}M
\eeq
boost invariance is also broken (assuming translations are unbroken). As such, there is an IHM at play and the Ward identities may be saturated without the need for a dilaton.  This begs the question,
can one write down a sensible  dilaton kinetic term if there is no particle condensate? 
The answer, as will be discussed below, is yes as long as the framon is included in the action.  So we see that the questions of the framon and the dilaton
are intimately connected. Thus the puzzle of the non-relativistic dilaton remains, and its resolution is tied to
the fate of the framon. 

As will be discussed below a resolution of the framon puzzle is closely related to the fact that
when space-time symmetries are broken, Goldstone bosons need  not be derivatively coupled.
To explore this possibility  we utilize the   coset construction which will allow us to generalize the criteria for non-derivatively coupled Goldstones given in \cite{VW}\footnote{We generalize \cite{VW} in two ways. \cite{VW} states that there can be no Goldstone associated with boost invariance due the non-vanishing commutator between boosts and the Hamiltonian. Here we show the need for the boost Goldstone and show that it couples non-derivatively. The coset methodology also allows us to consider relativistic generalizations.} . Furthermore, we will use the coset methodology to construct the theory of Fermi liquids with
the symmetry breaking pattern of type I/II framids as realized by a canonical Fermi liquid without/with nematic order. We will see that the resolution of the framon issue follows from the dynamics of the effective field theory.
By treating the Goldstone boson as a Lagrange multiplier we will generate  a set of constraints,
that are generalizations of the Landau conditions in canonical Fermi liquids, which when imposed,
lead to the proper symmetry realization.
We dub  this the ``Dynamical Inverse Higgs Mechanism" (DIHM), because the Goldstones are absent but not for
algebraic reasons.
 We will then go a step further
and discuss a more general symmetry breaking pattern where both boost invariance and  Schrodinger invariance are spontaneously broken, which we call a ``type-III framid''. In this case one might expect both a framon and a non-relativistic dilaton to arise.
We will see that again, they do not, but their absence greatly constrains the form of the effective field theory.
This analysis was recently used to prove that degenerate electrons interacting in  the unitary limit can not behave like a Fermi liquid \cite{us}
in the unbroken phase.
\subsection{The Paths to Symmetry Realization}

We see that there are three paths to space-time symmetry realization: No inverse Higgs constraints are applied and the system
retains one Goldstone for each broken generator. Some or all of the constraints are applied and we have a reduced number of Goldstones due to the existence of IHCs, or a Goldstone can be eliminated via the DIHM with or without the application of other inverse Higgs constraints. In this paper we will consider all three scenarios in the context of degenerate fermions. 
We will show two examples of DIHMs, one for boosts and the other for dilatations.


\section{Review Coset Construction}

A powerful method for generating actions with the appropriate non-linearly realized
broken symmetries was developed for internal symmetries by CCWZ\cite{Coleman:1969sm,Callan:1969sn} and later
generalized to space-time symmetries by Volkov and Ogievetsky \cite{Volkov,Ogievetsky}. We refer the reader to original
literature for details and here only rapidly review the salient points of this coset construction. The method uses the fact that  the Goldstones 
coordinatize the coset space $G/H$ where $G$ is the symmetry group of the microscopic
action and $H$ is the symmetry sub-group left unbroken by the vacuum. 
The vacuum manifold is parameterized by
\beq
U= e^{i \vec \pi \cdot \vec X}
\eeq
where $\vec \pi$ are the Goldstone fields and $\vec X$ the corresponding broken generators.
The unbroken generators will be denoted by $\vec T$.
This parameterization will  be generalized when we break space-time symmetries.
As discussed below, we may use $U$ to write down the most-general action
consistent with the symmetry breaking pattern, including terms where the Goldstone couples
to other gapless (non-Goldstone) fields in the theory. Notice that the coset construction seems
to imply that there must be at least one Goldstone boson. However, this need not be the case, as mentioned above.
It could very well be that we can construct an invariant action without the need for a Goldstone,  even without an inverse Higgs constraint.
We will show that if this is indeed possible then the coset construction is a  useful tool
in determining this {\it non-Goldstone} action. 

Once space-time symmetries are broken, the symmetry group is no longer compact.
As such the structure constants can not necessarily be fully antisymmetric \footnote{A consequence of this fact is that it is not longer true that $[T,X] =X$.} and consistency requires that  one generalize the vacuum parameterization to include
the unbroken translations ($\bar P^\mu$) \footnote{ When translations are broken by localized semi-classical objects (i.e. defects) the coordinate is lifted to the status of a dynamical variable see for instance \cite{wheel,mac,gomis}. } such that 
\begin{equation}
U=e^{i \bar P\cdot x}e^{i\pi \cdot X}.
\end{equation}
 The number of unbroken translations may be enhanced if there exist internal translational symmetries  as in the
 case of solids or fluids \cite{intsyms}. In such cases the direct product of the internal and space-time translations
 are broken to the diagonal subgroup by the solid. In this work we will not be considering such cases as
 we are interested in zero temperature ground states with delocalized particles.

The Maurer-Cartan (MC) form decomposes into a set of well defined geometrical objects,
\begin{equation}
\label{MC}
U^{-1} \partial_\mu U = E^A_\mu(\bar P_A+ \nabla_A \pi^a X^a + A^b_A T^b).
\end{equation}
The vierbein $E$ relates the global frame to the
transformed (acted upon by $G/H$) frame.
In this way, the covariant derivatives on the matter fields  in the local frame are written as
\beq
\label{mcovd}
\nabla_A \psi \equiv (E_A^\mu\partial_\mu+ i T^q A^q_A) \psi
\eeq
such that under a boost
\beq
\nabla_A \psi \rightarrow e^{\frac{i}{2}mv^2 t-i m \vec v \cdot \vec x}\nabla_A \psi.
\eeq

From (\ref{MC}) we can extract the vierbein, the covariant derivative of Goldstone fields $(\nabla\pi)$ and the Gauge fields $(A)$ and use these objects to construct our action which will be invariant under the full symmetry group $G$ by forming $H$ invariants.  For a complete discussion of the coset construction and its application to broken spacetime symmetries in multiple contexts, we refer the reader to \cite{wheel}.

\section{Non-Derivatively Coupled (NDC) Goldstone Bosons}
In \cite{VW} the criteria necessary to generate theories with non-derivatively coupled
Goldstones is given by
\beq
\label{crit}
[X_i, \vec P] \neq 0.
\eeq
where $X^i$ is a broken generator and $\vec P$ are the unbroken space-time translations.
The authors argue that  the  forward scattering matrix elements of broken generators $X$
formally diverge
\beq
\lim_{\vec k^\prime \rightarrow \vec k} \langle \vec k \mid X \mid \vec k^\prime \rangle \rightarrow \infty.
\eeq
 which compensates for the explicit factor of the Goldstone momentum in the coupling.
One may be concerned with the fact that $X$ is not a well defined operator at infinite volume,
and that the limiting procedure is not well defined. However, we will see below that the coset construction
supports  the authors claims and allows us to, trivially, generalize their criteria to relativistic systems. (\ref{crit}) is a necessary but not a sufficient criteria for the existence of a non-derivatively coupled Goldstones since we must also ensure that it can not be removed via an IHM. 

Within the coset formalism the search for non-derivative couplings starts with understanding
how the Goldstones couple to generic matter fields. As such, we need to determine under what
conditions a Goldstone arises in the vierbein or connection without any derivatives acting upon it.
Thus  a necessary condition for non-derivative coupling is the generalization of $(\ref{crit})$, i.e.
\beq
\label{ours}
[\bar P^\mu,X] \neq 0.
\eeq
Note the distinction between this criteria and (\ref{crit}). First (\ref{ours}) only  involves the unbroken canonical spatial translations $\bar P$ which 
can differ from $P$, not only because of the zero component, but more generally if there are internal translational symmetries. 
This is however, a distinction without a  difference because internal and space-time symmetries 
commute. But an important distinction between (\ref{crit}) and (\ref{ours}) is the fact that (\ref{ours}) 
allows for the non-commutation with the Hamiltonian as being a criteria for NDC
Goldstones. As a matter of fact, this explains the NDC nature of the dilaton (both relativistic as
    well as non-relativistic\footnote{The non-relativistic case being of particular importance below.}).
Also we will see that whether or not $G$ is the Poincare or Galilean group is of no consequence as far
the the criteria for non-derivative coupling is concerned. 

To see that (\ref{ours}) is a sufficient criteria for NDC, assuming the Goldstone boson associated with $X$ is not
removed by the inverse Higgs mechanism, we note that the veirbein will contribute to the measure via
\beq
S= \int d^4x \sqrt{E^2}.....
\eeq
 so that as long
as the determinant of the vierbein contains a term linear in the Goldstone\footnote{That $E$ contains term linear in the Goldstone follows from the fact that  the Goldstone acts as the transformation parameter.} , there will be a NDC to matter fields.
From (\ref{MC}) we can see that if $[\bar P, X] \sim \bar P$ then the Goldstone associated with $X$ will arise in $E$.
However, the Goldstone will often be absent from the volume factor as in the case of broken boosts or rotations.
Thus the first NDC will come from the covariantization of the derivatives $E^\mu_a(\pi) \partial_\mu$.
Alternatively if  $[\bar P,X] \sim T$, then the  Goldstone will show up in the connection, in which case the
NDC will arise from the covariant derivative acting on the matter fields.

Finally, note that if $G$ is the Galilean group then due to relation the Eq.~(\ref{U1}) if the $U(1)$ particle number is unbroken, then the boost Goldstone will be associated with the connection.
Whereas if $G$ is the Poincar\'e group then the boost will be in the vierbein. But in either case framid will
be non-derivatively coupled. 

\section{Framids}
\subsection{Non-Relativistic Framids}
The type I framid as defined in \cite{framids} is a  system where boost invariance is spontaneously broken, but all other
space time symmetries are intact.
The coset construction only cares about the symmetry breaking pattern and not the definite choice of the
order parameter. As was emphasized in \cite{gapped} the choice of order parameters can affect how
the symmetry is realized if there exist gapped Goldstones (assuming the gap size is hierarchically small
compared to the cut-off). In particular the  representation of the  order parameter(s) will determine whether or not
the the inverse Higgs conditions (\ref{IH}) leads to a redundancy or a gap. However, here we are only 
interested in the truly gapless modes, so in this respect the order parameter  will be irrelevant. Nonetheless, we are interested in a certain class of order parameter, i.e. those whose commutator with boost generators have a non-vanishing vacuum expectation value (e.g. the momentum density). This class of order parameters yield  Goldstones which  are collective excitations. Whereby a ``collective excitation'' we mean 
a quasi-particle  pole (or resonance)  which exists as a consequence of the fact that the vacuum is not annihilated by some
conserved charge. Put another way, the modes are excitations of the material
responsible for the breaking of boost invariance. This definition sets apart say the pion in QCD from the plasmon in a metal.

Cases where the framid are not collective modes correspond to speculative theories beyond the
standard model of particle physics and Relativity, such as Einstein-Aether theory \cite{theo}, where a four vector gets a time-like expectation value.
\beq
\langle A_\mu \rangle = n_\mu. 
\eeq
The resulting theory contains 3 Goldstone modes corresponding to the framons \cite{riccardo}. The lack of the evidence for
a Goldstone arising in Einstein-Aether theory allows us  to place bounds on the couplings (see for instance \cite{theobounds}).
However, we know that condensed matter systems break boost, and if the symmetry breaking pattern is such that  there are no IHC around to eliminate the framids from the spectrum it is incumbent upon us to determine their fate.




It is tempting to disregard  boost Goldstones since the associated generator does not  commute with 
the Hamiltonian and hence there is no flat direction.
However,  the existence of the relativistic dilaton immediately dispels this notion. 
Furthermore,  the inclusion of the framid
into the coset parameterization is {\it necessary} for consistency. Moreover, according to the criteria for NDC (\ref{ours}) we should expect  the coupling to the framid
to be at least marginal.


To manifest framids in the laboratory we need systems which break boosts yet whose ground
state does not break any symmetry which would lead to an inverse Higgs constraint. Thus we may eliminate
electrons moving in a crystal background as well as super(fluids/conductors) from the list of possibilities. It would seem that we are
relegated to degenerate electrons in the unbroken phase.
 One might be concerned that the Kohn and Luttinger  \cite{kohn} effect ensures that {\it all} Fermi liquids superconduct, even if the coupling function is repulsive in all channels in the UV. However, 
 all we really need to manifest a framon is for there to be a temperature window between the
 boost symmetry breaking scale ($E_F$), and the critical temperature $T_c$.
 For a Fermi liquid the critical temperature 
 scales as
 \beq
 T_c \sim \Lambda_\star\ll E_F
 \eeq
 where $\Lambda_\star$ is the strong coupling scale which is typically exponentially suppressed. Thus there is a range of temperatures where the framid should contribute to the
 heat capacity.
This is as opposed to the bosonic case where the critical temperature is set by the number density
\beq
T_C \sim n^{-1/3}.
\eeq
and the boost symmetry breaking scale is of the same order. 


Thus we have narrowed our search  for framons to degenerate Fermi gases  whose phenomenology  certainly shows no signs of non-derivatively coupled Goldstone. One might be tempted to interpret zero sound as the boost Goldstone,
however,  the interaction between
electrons due to zero sound exchange vanishes in the forward scattering limit. 
\subsection{Coset Construction of Fermi Liquid EFT with Rotational Symmetry: Type I Framid}
We begin our investigation by building the coset construction for type I framids (i.e. systems with broken boosts but unbroken rotations).
We  consider the case of broken Galilean invariance, as the relativistic case will follow in a
similar manner.

The vacuum manifold is  parameterized
by 
\begin{equation}
U= e^{i P \cdot x }e^{-i \vec K \cdot \vec \eta(x)}
\end{equation}
Calculating the MC-form, we can extract the vierbein
\begin{equation}
E_0^0= 1~~~E_i^j = \delta_i^j, ~~~E_0^i= \eta^i, E_i^0=0.
\end{equation}
The gauge field is given by
\begin{equation}
A_i= -\eta_i, A_0= -\frac{1}{2}\vec \eta^2
\end{equation}
and the covariant derivatives of the framids are (up to lowest order in fields and derivatives )
\begin{equation}
\nabla_0 \eta^i= \dot \eta^i~~~\nabla_i \eta^j= \partial_i \eta^j.
\end{equation}
The free action  for the Goldstone follows by writing down all terms which are invariant under the linearly realized $H$ symmetry
\begin{equation}
\label{framonKE}
S =\int d^dx dt \Big(
 \frac{1}{2}
  \dot \eta_i^{ 2}
- \frac{1}{2}u_T^2 ( \partial_i  \eta_j)^2-\frac{1}{2}u_{L}^2 (\partial \cdot \eta)^2\Big)
\end{equation} 
Following eq. (\ref{mcovd}), the coupling for the Goldstone to matter fields via the covariant derivative
 is given by
\begin{eqnarray}
\label{fermiquad}
S_{0}
&=&\int d^dx dt~\psi^\dagger\left[ i(\partial_0+\eta^{ i}\partial_i) +  \frac{1}{2}m {\vec \eta^{2}} + \varepsilon(i \partial_i +m\eta_i )\right] \psi 
\end{eqnarray}
where  $\varepsilon$ is the unknown dispersion relation that is fixed by the dynamics. Due to the central extension of the Galilean algebra, the fermion under a boost transformation with velocity $\vec{v}$ transforms as
\begin{equation}
\label{boost}
\psi(x,t) \rightarrow e^{\frac{i}{2}m \vec v^2t- im \vec v \cdot \vec x}\psi(x,t).
\end{equation}
while the Goldstone field $\eta$ undergoes a shift
\begin{equation}
\label{boostshift}
\vec \eta \rightarrow \vec \eta + \vec v.
\end{equation}
The $\eta^2$ term will be sub-leading and not play role in the remainder of our discussion. 

As in the standard EFT description of Fermi liquids \cite{shankar,polchinski}
the quasi-particle self interaction is  most conveniently written  in momentum space
\begin{equation}
\label{qpint}
S_{int}= \prod_{i,a}\int d^dk_i dt ~g(\vec k_i+m\vec \eta)\psi^\dagger_{k_1}(t) \psi_{k_2}(t) \psi^\dagger_{k_3}(t) \psi_{k_4}(t) \delta^d(\sum_i k_i)
\end{equation}
Higher order polynomials in the matter field $\psi$ are technically irrelevant (see below). 
$g$ is the coupling function which now formally depends upon the framon. The assumption of spherical symmetry implies
$g$ is a scalar.
 Notice that the $\eta$ is non-derivatively coupled, as expected from our considerations of the algebra, which  can lead to non Fermi liquid behavior. Given that ${\rm He}^3$, e.g., is well described by Fermi liquid theory, the framid must somehow decouple, yet it must do so in
such  a way that the theory remains boost invariant.
\subsection{Multiple Realizations Of Broken Symmetry}
Before moving onto further discussion about the framids in Fermi liquids, we want to highlight a subtle point about non-linear realizations of broken symmetries, which is that the same symmetry breaking pattern can lead to contrasting physical theories with very different particle content. This usually happens when there are two different order parameters. However, below we show that even with same order parameter we can have two different realizations of the symmetry. An example of this is the case of a massive complex scalar particle $(\phi)$ coupled to  gauge fields.
To power count this theory it is useful to introduce the notion of a field label as was introduced in Heavy Quark Effective Theory (HQET) where one is interested in the dynamics of a massive source which interacts with light gauge fields carrying momenta much less than the quark mass. The label is introduced by defining a re-phased field
\beq
\phi(x)= \sum_v e^{i m v \cdot x} h_v(x),
\eeq
such that  $v$ defines a superselection sector\cite{howard}. Derivatives acting on $h_v(x)$ scale as ``residual momenta'' ($k$)
which obey $k \ll m$. The vacuum of the system, labeled by $v$, breaks boost invariance
 and so we expect that framid should exist as an independent degree of freedom. Typically, the Goldstone modes are associated with collective excitations of a system which are clearly absent as the choice of vacuum is not dynamical. 
 Nonetheless the boost invariance must be non-linearly realized. 

Using the covariant derivatives derived in the previous section, we can write down the most general action for $\phi$ which is invariant under translations and rotations,
\bea
\label{scalarHQET}
\mathcal{L}_{\phi}&=&\frac{i}{2}\left(\phi^{\dagger}(t,\vec{x})(\partial_t + \vec \eta \cdot \vec \partial +\frac{i}{2}m \vec \eta^2) \phi(t,\vec{x})-[(\partial_t + \vec \eta \cdot \vec \partial-\frac{i}{2}m \vec \eta^2)\phi^{\dagger}(t,\vec{x})]\phi(t,\vec{x})\right) \nn \\
         &+& \frac{c_{1}}{2m}( (\vec \partial +i m \vec \eta)\phi^{\dagger}(t,\vec{x}))(\vec \partial -i m \vec \eta) \phi(t,\vec{x}).
\eea


If we choose $c_1=1$ then $\eta$ decouples from $\phi$ and we get the standard non-relativistic kinetic term for a free particle.  Had we started with a theory without the $\eta$, then $c_{1}$ can be fixed by requiring the theory to obey Galilean algebra, in particular by satisfying the commutator $[H,K_{i}]=iP_i$.  $c_1$ can equally well be fixed by Reparametrization Invariance (RPI) \cite{RPI}, which is related to the freedom in splitting the heavy quark momentum into a large and small piece (more on this below). However we can leave $c_1$ to be completely arbitrary and keep $\eta$ in the spectrum and the theory will still respect all the symmetries. The two theories (with and without $\eta$) are completely different and we have no reason to believe they will lead to same physics in the IR and yet they have the same symmetry breaking pattern and the same order parameter (local momentum density). Thus there are multiple ways of realizing the  boost symmetry. While it would seem that this is a rather trivial example, we note that only difference between the HQET ground
state and that of a Fermi liquid lies in the change in the number density from one to Avogadro's number.

 \subsection{Power Counting}
To determine the possible symmetry realization in a Fermi liquid, we must first discuss the systematics of the relevant  EFT whose action is given by ($\ref{fermiquad}$).
The matter fields (which we will call electrons from here on)  are effectively expanded around the Fermi surface, by removing the large energy and
momentum components via the redefinition
\beq
\psi(x)= \sum_{\theta} e^{i \varepsilon(k_F)t}e^{-i  {\vec {\bf k}}(\theta) \cdot \vec x}\psi_{{ \vec {\bf k}} (\theta)}(x),
\eeq
the assumption of rotational invariance implying that the magnitude of $\mid \! \vec {\bf k}(\theta)\!\mid =k_F$.
The field label ${{ \vec {\bf k}} (\theta)}$ is the large momenta around which we expand. 
As opposed to the HQET case, here the bins are dynamical and there is no super-selection rule.
This case is more akin to NRQCD \cite{vNRQCD} where the labels change due to Coulomb exchange.
Notice that there is a sum over the labels as opposed to an integral, this illustrates the fact that we have effectively tessellated the Fermi surfaces into ``bins". The size of each bin will scale as $\lambda\sim E/E_F$. The fact that theory should not depend upon
the bin size imposes constraints on the action. That is, we should be able deform the momentum around any fixed value we wish, by an amount scaling as $\lambda$, and the theory should be invariant. This re-parameterization invariance (RPI) \cite{RPI} implies that the action can only be a function of $\vec k_F +\vec \partial$.
In general RPI generates relations between leading order and sub-leading Wilson coefficients.
It is convenient for power counting purposes to introduce a label operator ${\bf \cal P}$ \cite{label}
such that 
\beq
{\bf \cal \vec P} \psi_{{ \vec {\bf k}} (\theta)}(x)={{ \vec {\bf k}} (\theta)} \psi_{{ \vec {\bf k}} (\theta)}(x).
\eeq
Full theory derivatives then decompose into the RPI invariant combination ${\bf \cal \vec P}+i\vec \partial$.
In this way we may drop the exponential factors as long as we assume label momentum
conservation at each vertex. The action becomes
\begin{eqnarray}
\label{action1}
S_0
&=&\sum_{{ \vec {\bf k}} (\theta)}\int d^dx dt~\psi_{{ \vec {\bf k}} (\theta)}^\dagger(x)\left[i\partial_0-\vec \eta(x)\cdot ({\bf \cal \vec P}+i\vec \partial) +  \frac{1}{2} m{\vec \eta (x)}^2 + \varepsilon({\bf \cal \vec P}+i \vec \partial +m\vec \eta (x))-\varepsilon(k_F)\right] \psi_{{ \vec {\bf k}} (\theta)}(x) \nn \\
\end{eqnarray}
Notice that the interaction with $\eta$ does not change the quasi-particle label. The reasons for this will
be discussed below once we have fixed the power counting systematics. Under a boost the labels are left
invariant but the residual momentum shifts. Furthermore under a boost the time derivative transforms as
\beq
i\partial_0 \rightarrow i\partial_0+\vec v \cdot \cal \vec P.
\eeq

 \begin{figure}
    \centering
    \includegraphics[width=12cm]{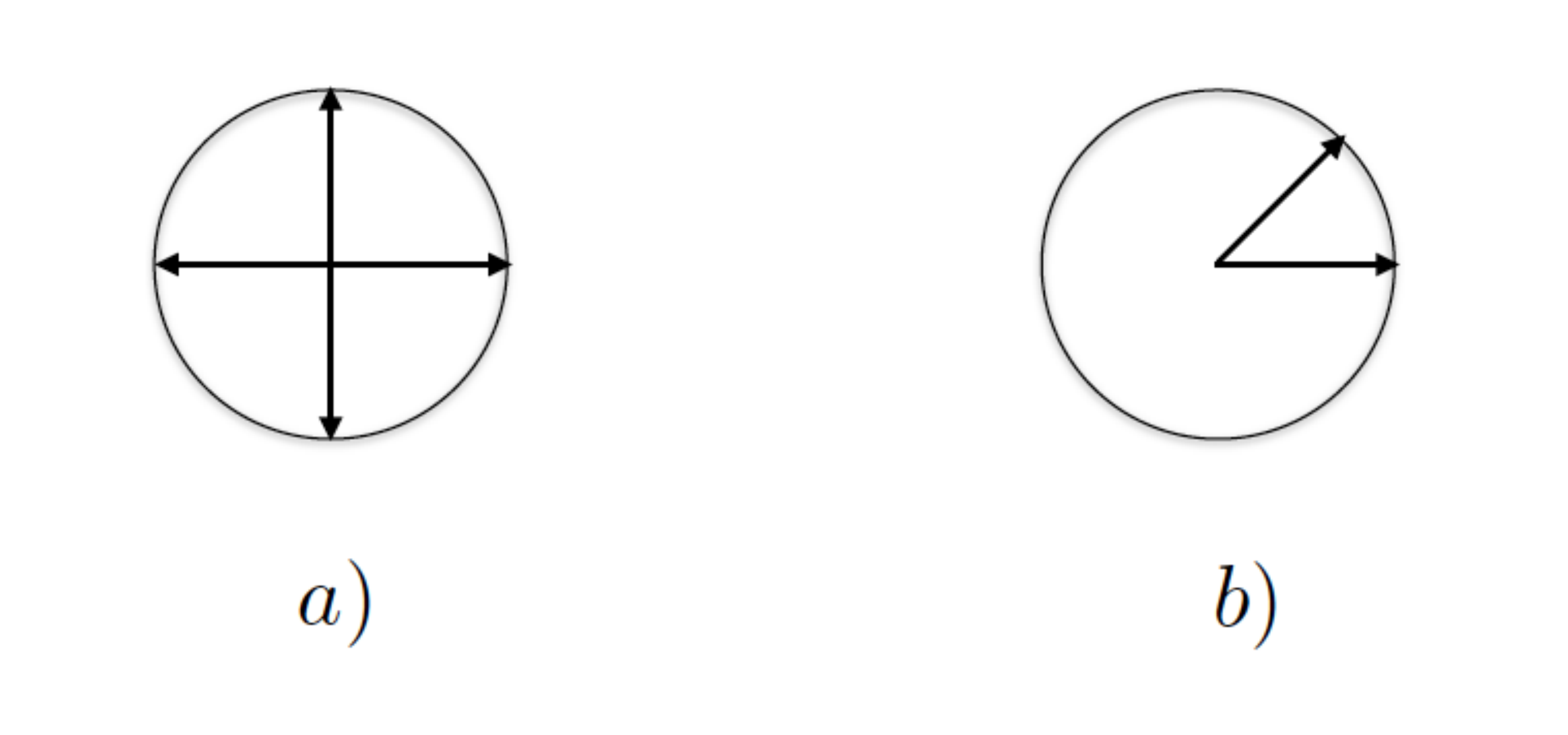}
\caption{Allowed kinematic configuration  for quasi-particle scattering.  Diagram (a) is the BCS back to back configuration which leads to Cooper pair condensation. (b) Forward scattering, in which the final state momenta lie on top of the
initial state momenta.}
\label{DIS}
\end{figure}

\begin{figure}
	\centering
	\includegraphics[width=10cm]{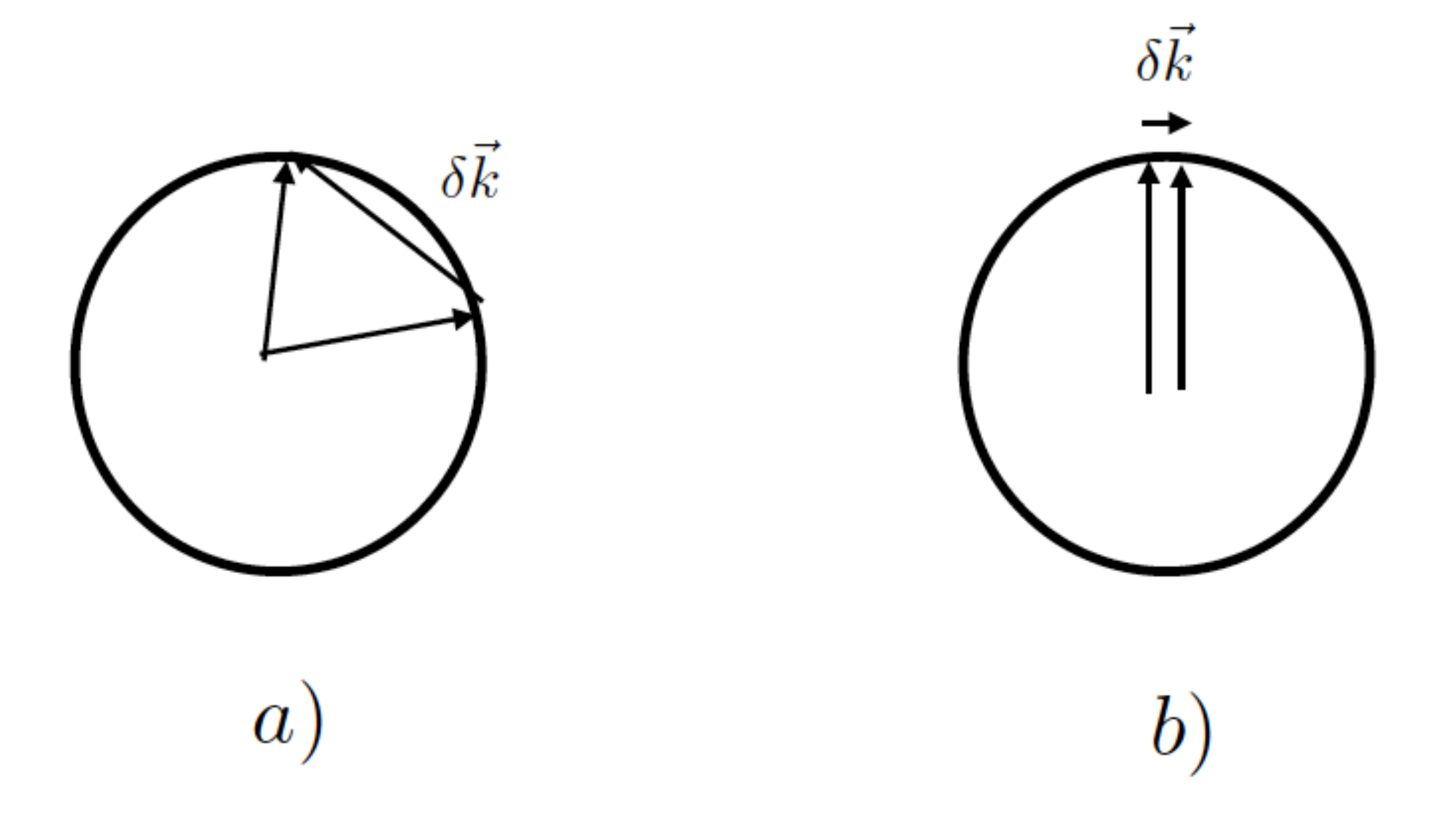}
	\caption{Allowed kinematic configuration  for framid-quasi-particle scattering.  Diagram (a) involves
		an off-shell framid, which can be integrated out. (b) shows the interaction with a soft framid leading to near forward scattering.}
	\label{DIS}
\end{figure}

\subsubsection{Review of EFT of Fermi Liquids Scalings}
We first review the EFT of Fermi liquids and its power counting  (for details see \cite{benfatto,shankar,polchinski}).
In the EFT, the power counting is such that the momenta perpendicular to the Fermi surface ($k_{\perp}$) scale as
$\lambda \sim E/\Lambda$, where the theories' breakdown scale is $\Lambda \sim E_F$. 
With this scaling, the most relevant terms in the action come from expanding the energy and coupling function
around the Fermi surface and keeping the leading term in $k_\perp$.

 The scaling of the electron field 
\beq
\psi(\vec k ,t) \sim \lambda^{-1/2},
\eeq
 follows from the equal time commutator
\beq
\{ \psi(\vec k,t=0),\psi^\dagger(\vec p,t=0)\} \sim \delta( k_\perp -  p_\perp)\delta^{d-1}( k_\parallel -  p_\parallel) \sim 1/\lambda
\eeq
since $k_\parallel $ does not scale.
Thus ignoring the framon for the moment,
the leading order action is given by \footnote{We are ignoring spin as it will not play a role in our discussion.}
\begin{equation}
\label{qpint}
S_{int}=\sum_{{ \vec {\bf k}}} \int d^dl dt~\psi_{{ \vec {\bf k}} }^\dagger(t,l) ( i\partial_0  +  \vec{ l}_\perp \cdot \vec v_F)\psi_{-{ \vec {\bf k}} }(t,-l)
+\sum_{{ \vec {\bf k}_i}}\int d^dl_i dt ~\frac{g({ \vec {\bf k}_i})}{2}\psi^\dagger_{\bf k_1}(t,l_{1}) \psi_{\bf k_2}(t,l_{2}) \psi^\dagger_{\bf k_3}(t,l_{3}) \psi_{\bf k_4}(t,l_{4}) \delta^d(\sum_i l_i).
\end{equation}
The Fermi velocity defined as $ \vec v_F=\frac{\partial \varepsilon}{\partial k^i_\perp }\!\!\mid_{k_F}$
 is constant on a spherically symmetric Fermi surface. In the last term there is a Kronecker delta for the label momenta
 that is implied. The residual momenta scale as $l_\perp \sim \lambda$ and $l_{\parallel}\sim 1$. 
 The latter scaling might seem odd given that it is a residual momentum. However, $l_\parallel$ scales as the bin size,
 which does not play a role for Fermi surfaces which are featureless
 Another way of saying this is that the $l_\parallel$ integral
 can be absorbed into the label sum.

Naively the interaction terms looks  irrelevant because the delta function scale as $\lambda^0$ for generic
kinematic configurations so that, once the scaling of the measure is taken into account ($\sim \lambda^3$), the operator will scale like $\lambda$.
However, there are two configurations for which one of the delta function will scale as $1/\lambda$: The BCS configuration (back to back incoming momenta) and
forward scattering. These two configurations are shown in figure (1), where it can be seen that these are the only two 
possible configurations that allow for momentum conservation that keep all momentum within $\lambda$ of the
Fermi surface.

It is convenient to decompose the BCS coupling into partial waves $g_l$. A one loop calculation (which is exact) of the beta function shows that $g_l$  are 
  either marginally relevant/irrelevant for attractive/repulsive UV initial data.
 The forward scattering coupling does not run, but plays an important role in the IR nonetheless.
 Interestingly, below we will show  that  Galilean invariance is sufficient to prove that  the forward scattering and BCS kinematics
 are the only possible marginal/relevant interactions.  This result follows without the need to consider the effects of the special
 kinematics on the power counting of the four Fermi operator.

\subsubsection{ Power Counting  in the Coset Construction}
Let us now derive the power counting from the coset construction.
%
We begin with the kinematics of the framid interactions.
The two allowed scattering configurations are shown in figure (2). Figure (a) shows the interaction of a quasi-particle
with a framid that is far off its mass shell in the sense that $E\ll k$, in the EFT language this would be called a
``potential framid" and can be integrated out. Thus these interactions are swept, along with those of the phonon and screened electromagnetic interactions, into a non-local coupling. Note that the potential is effectively  local because
the labels on the incoming and outgoing quasiparticles can not be the same and hence it is analytic in (the small) residual momenta\footnote{In this sense it is better to think of the $1/k^2$ in the interaction as a Wilson coefficient.}.
Figure (b) shows the interaction with an on-shell framon whose momentum is necessarily soft $k \sim E\ll E_F.$
If we define our power counting parameter as $\lambda \sim E/E_F$, then we only know that $k\sim \lambda^n$,
where $n$ is yet to be determined. However, symmetries fix $n$ as the covariant derivative must scale homogeneously in $\lambda$
for the theory to be boost invariant.  That is, $\eta$ must scale in the same way as the residual momentum of order $\lambda$,  so that  $\eta \sim \lambda$.
Given that $\partial \sim \lambda^n$ we can fix $n$ by considering the canonical commutator
\beq
[\eta^i(x), \dot \eta^j(0)]\sim \lambda^{n+2} \sim \delta^d(x)\delta^{ij} \sim \lambda^{dn}
\eeq 
thus  $n = \frac{2}{d-1}$.  
Thus we see that in two spatial dimensions $k \sim \lambda^2$ and the framons can not change the (residual) momentum  of the quasi-particles
and only their zero mode is relevant. This however is not the case in three dimensions where the framon
carries off residual momentum $k \sim \lambda$.

Expanding the action (\ref{action1})
\begin{eqnarray}
S_0&=&
\sum_{{ \vec {\bf k}} (\theta)}\int d^dx dt~\psi_{{ \vec {\bf k}} (\theta)}^\dagger(x)\left[ i\partial_0-\vec \eta(x)\cdot { \vec {\bf k}} (\theta) +   (i\vec \partial +m\vec \eta(x)) \cdot \frac{\partial \varepsilon}{\partial k}\right] \psi_{ \vec {\bf k}(\theta)} (x)+... \\
&=& \sum_{{ \vec {\bf k}} (\theta)}\int d^dx dt~\psi_{{ \vec {\bf k}} (\theta)}^\dagger(x)\left[ i\partial_0 - \vec \eta(x)\cdot { \vec v_F} (\theta)(m-m^\star) + i\vec v_F(\theta) \cdot \vec \partial\right] \psi_{ \vec {\bf k}(\theta)} (x)+...
\nn \\ \end{eqnarray}
where $m^\star$ is the effective mass defined by $ \frac{\partial \varepsilon}{\partial \vec k}= \vec v_F= \frac{\vec k_F}{m^\star}. $
In two dimensions we must multipole expand the framon field 
to preserve manifest power counting \cite{GR}, which leaves only the coupling to the
framon zero mode.
The leading order action is given by
\begin{eqnarray}
S_0^{d=2}&=& \sum_{{ \vec {\bf k}} (\theta)}\int d^2x dt~\psi_{{ \vec {\bf k}} (\theta)}^\dagger(x)\left[ i\partial_0-\vec \eta(0)\cdot { \vec v_F} (\theta)(m-m^\star) + i\vec v_F(\theta) \cdot \vec \partial\right] \psi_{ \vec {\bf k}(\theta)} (x)+...
\nn \\ \end{eqnarray}

From here on to simplify the notation we will be dropping the label sum and the bold font for labels as all momenta
unless stated otherwise will be labels.

Before we move on to determine the consequences of the multipole expansion let us pause to
clarify this unusual scaling.
Typically in an EFT the scaling of the fields follows from the scaling of the momenta not the other way around as in this case.
Indeed, it would be useful to understand what happens to loops with momenta scaling as  $\lambda$ and not $\lambda^2$.
However, symmetries  forbid such contributions and it must be that if we do not multipole expand
the framon interaction, that power counting and boost invariance are incompatible.
Thus we see that in two spatial dimensions, the symmetries can not be realized via a Goldstone as the
framon equations of motion allow us to eliminate it from the theory, as will be discussed below.
In three spatial dimensions this conclusion does not follow.

\subsubsection{The Framid as Lagrange Multiplier and the Landau Relation}
\begin{figure}[t]
    \centering
    \includegraphics[width=15cm]{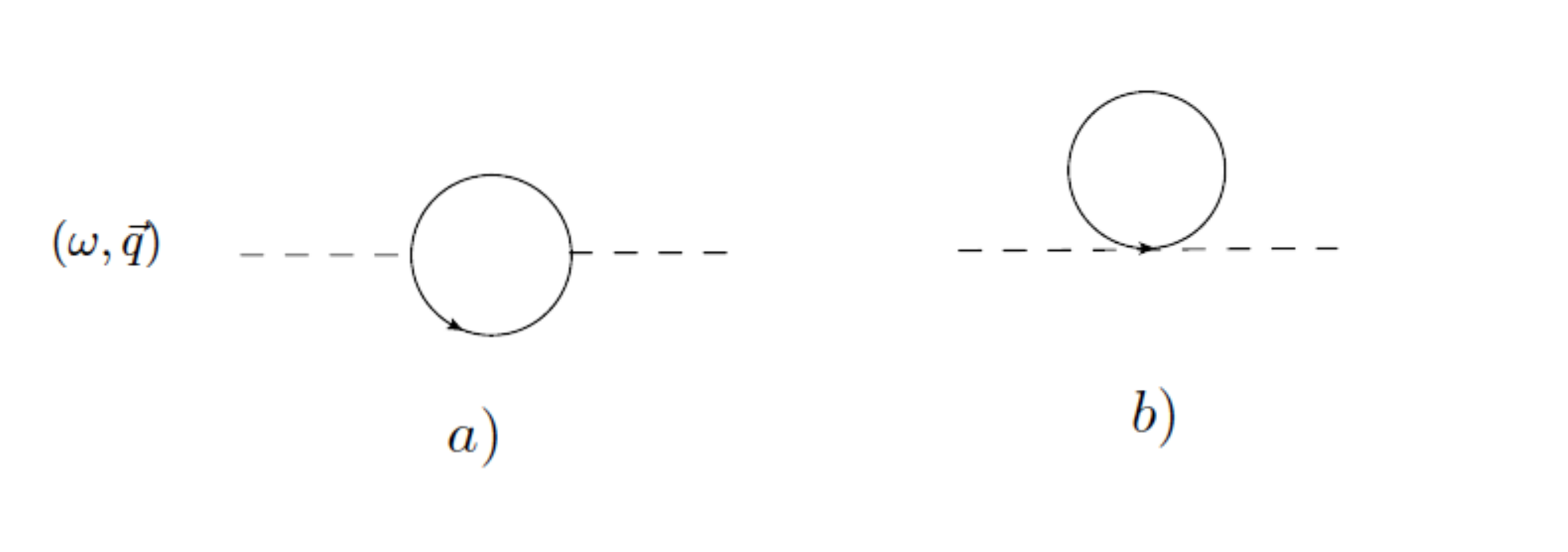}
\caption{Diagram a)  could contribute to wave function renormalization whereas both a) and b) could contribute to a mass.
At zero external momentum the two diagrams cancel as dictated by boost invariance.}
\label{Zeta}
\end{figure}

Let us  consider the ramifications of the multipole expansion of the framon  in two dimensions. \footnote{Even though our arguments in this section are strictly valid only for d=2, we keep d arbitrary to generalize it latter to d=3.}
 Since the kinetic piece of the Framon action vanishes for  the constant zero mode $\eta$
plays the role of a Lagrange multiplier.

Expanding the action for the four-Fermi interaction term leads to the coupling
\begin{equation}
\label{qpint}
S_{int}= \prod_{i}\int d^dk_i dt \sum_j \frac{m}{2}\vec \eta\cdot \frac{\partial g(k^j)}{\partial \vec k^j}\psi^\dagger_{k_4}(t) \psi^\dagger_{k_3}(t) \psi_{k_2}(t) \psi_{k_1}(t) \delta^d(\sum_i k_i).
\end{equation}
Using the equations of motion for  $\eta$ gives the operator constraint $O^B_i=0$ where 
\begin{equation}
\label{GL}
O^B_i=\int \frac{d^dp}{(2\pi)^d}~\psi^{\dagger}_{p}(t)\left(p_{i}-m \frac{\partial \varepsilon_p}{\partial p_i} \right)\psi_{p}(t)
-\frac{m}{2}\int\prod_{a=1}^4\frac{d^{d}p_{a}}{(2\pi)^{d}} \delta^{(d)}(\sum_{i} p_{i})\Big(\sum_{i}\frac{\partial{g(p_{a})}}{\partial{p_{i,a}}}\Big)\psi^\dagger_{p_4}(t)\psi^{\dagger}_{p_3}(t)\psi_{p_2}(t)\psi_{p_1}(t).
\end{equation}
This is a strong operator constraint  both  technically and colloquially.
Notice that the constraint is non-local in the sense that it is integrated. 
This is crucial, as the constraint is a function  of the Noether charges. Indeed, current
algebra imposes this same constraint $O^B_i=0$ as shown in the appendix where we also derive
the relativistic generalization of this constraint.

The power counting of the terms in this constraint deserve attention. 
The first two terms scale as $\lambda^0$ while the last
term naively scales as $\lambda^2$  as the measure scales as $\lambda^4$.
Recall  that at this point we have not made any assumption about special
kinematics so the delta function does not scale. We are trying to derive the fact that the only relevant
couplings have these special kinematics.
Thus we might naively think that we can drop the quartic term in the constraint. In general this is true, but there is
an exception as we now explain. We begin by noticing that the quartic term is time dependent while the quadratic terms (being
conserved charges) are not. 
Thus it would seem that the last term must  vanish (to the order we are working). However, if we insert the quartic
term in a two point function the time dependence will cancel.


Consider taking matrix of element of (\ref{GL}) in a 1-particle state with momentum $\vec k$  ($|\vec{k}|=k_{F}$) \begin{equation}
\label{landau}
k_{i}=m\frac{\partial\varepsilon(k)}{\partial{k_{i}}}+\frac{2m}{(2\pi)^{d}}\int d^d{p}\Big(\frac{\partial{g(k,p,p,k)}}{\partial{p_{i}}}+\frac{\partial{g(k,p,p,k)}}{\partial{k_{i}}}\Big)\theta(p_F-p)
\end{equation} 
We can now see why that the interaction term is  enhanced because the radial integral, naively scaling as $\lambda$ is
actuality scaling as order one. This is a consequence of the power divergence of the integral. Such mixing of orders
is commonplace in effective field theories when a cut-off regulator is used. However, here the cut-off (the radius of the Fermi surface)
is physical\footnote{In canonical EFT's one uses dimensional regularization exactly to avoid this mixing issue which complicates
the power counting.}.

We re-write this result in the form
 \begin{equation}
\label{Mtxele}
k_{i}=m\frac{\partial\varepsilon_{k}}{\partial{k_{i}}}+2m\frac{\partial}{\partial{k_{i}}}\int \frac{d^d{p}}{(2\pi)^d}~\theta(p_{F}-p)g(p,k)+2m\int \frac{d^{d}p}{(2\pi)^d}~g(p,k)\delta(\varepsilon_{F}-\varepsilon)\frac{\partial\varepsilon_{p}}{\partial{p_{i}}}.
\end{equation} 
The second term on the RHS vanishes by spherical symmetry.


Next using the assumption of rotational invariance, and expanding the coupling function in Legendre polynomials, $g(\theta)=\sum_{l}g_{l}P_{l}(\cos{\theta})$\footnote{We take d=2 for sake of simplicity but the results are valid for arbitrary d.}
 we get
\begin{equation}
\label{Landau1}
\frac{k_{F}}{m}=v_{F} +\frac{2 p_F}{(2\pi)^2}\int d\theta~\cos{\theta}\sum_{l}g_{l}P_{l}(\cos{\theta}).
\end{equation}

Using this result we get the famous Landau relation \cite{Landau} for a Fermi Liquid
\begin{equation}
\label{Landau2}
\frac{m^
\star}{m}=1+\frac{1}{3} \frac{2 m^\star}{(2\pi)^2} g_1
\end{equation}
Notice that at this point it is not clear that this result will hold to all orders in perturbation theory.

It is interesting to ask whether or not more information can be extracted from the constraint by considering
a two body state. However, as is seen by inspection the insertion of the constraint operator  on external lines will
will automatically be satisfied once the Landau condition is imposed, and furthermore the insertion of the quartic function
into a four point amplitude will be  suppressed since there is no power divergence that can enhance its scaling.

We can glean more information from the Landau criteria by utilizing the fact that the equation is
RG invariant. Differentiating it with respect to the RG scale implies that the beta function vanishes.
For generic momenta the four point one loop interaction diverges logarithmically. 
To avoid this conclusion
we  impose a kinematic constraint
to suppress the one loop result. 
If we consider the forward scattering interaction, 
\beq
S_F = \int d^d x dt\sum_{{\bf k}(\theta_i)} g(\theta_1,\theta_2)\psi_{{\bf k}(\theta_1)}^\dagger(x,t) \psi_{{\bf k}(\theta_1)} (x,t)\psi_{{\bf k}(\theta_2)}^\dagger(x,t) \psi_{{\bf k}(\theta_2)}(x,t)
\eeq 
then the one loop result {\it vanishes} since the constraints imply that the loop involves no sum over
the large label bins leading to a power suppressed result.

It would seem that we have ruled out the possibility of a BCS interaction which has a non-vanishing beta function at one loop. However, this is not the case
as such an interaction would not contribute to the Landau relation since the tadpole diagram vanishes for the BCS interaction.

Thus we have reached the conclusion that the only allowed interactions are BCS and forward scattering.
We are not claiming that this is a rigorous proof since we have assumed that the only sensible coupling
with vanishing beta function is forward scattering. Furthermore, our argument regarding the acceptability of the BCS
coupling is based on the fact that our arguments allow for any coupling which leads to a vanishing tadpole (with no associated counter-term).
It is possible that there are other allowed kinematic configurations, however, assuming a featureless Fermi surface \footnote{In cases where the Fermi surface is singular there are other relevant interactions whose self contractions would vanish \cite{VH} algebraically.} we have not been able to find
any sensible  examples.

Recall at this point the result in the section only hold at one loop. However, now that we have restricted our interactions
to BCS and forward scattering we know that that the Landau relation holds to all orders. This well known result follows
from the fact that tadpole corrections to the one loop insertion of the constraint are pure counter-term and vanish.

Finally recall that this result assumed that the framon acts as a Lagrange multiplier. However, this was only forced upon us
in two dimensions. In three dimensions, there is the logical possibility that the framon remains in the spectrum and there is
no DIHM at play. This will be discussed below when we list the possible paths to symmetry realization in three dimensions.




\section{Fermi Liquid with broken rotational invariance }
Let us now consider the case where the rotational symmetry is broken by the Fermi surface (the typeII framid). We work in two spatial
dimensions for the sake of simplicity. 
Again, to avoid an algebraic inverse Higgs constraints, we assume that the $U(1)$ particle number
is unbroken.
 The vacuum will be parameterized by 
\beq
U(\vec \eta, \Theta,x)= e^{i P \cdot x} e^{-i \vec K \cdot \vec \eta(x)}e^{-i L \Theta(x)}
\eeq
The rotational Goldstone boson ($\Theta$) is called the ``angulon" has been studied
in the context of  electronic systems \cite{OKF} as well as in neutron stars \cite{NS}, although to our knowledge
its non-linear self interactions have not been previously derived.

Calculating the MC-form we may extract the vierbein
\begin{equation}
E_0^0= 1~~~E_i^j = R_i^j(\Theta), ~~~E_0^i= R^{ij}(\Theta)\eta^j, E_i^0=0.
\end{equation}
where $R(\Theta)$ is the two dimensional rotation matrix.
The gauge fields  are
\beq
A_i=-R_{ij}(\Theta) \eta_j,~~~A_0=-\frac{1}{2} \vec \eta^2.
\eeq
 the covariant derivatives of the angulons are
\begin{equation}
\nabla_0 \Theta= \dot \Theta_i+\vec \eta \cdot \vec \partial \Theta,~~~\nabla_i \Theta= R_{ij}(\Theta)\partial_j \Theta= \partial_i \Theta+ \epsilon_{ij}\Theta\partial_j \Theta+....
\end{equation}
The quadratic piece of the quasiparticle action is given by
\begin{eqnarray}
\label{Spsi}
S_{\psi}
&=&\int d^dx dt~\psi^\dagger\left[ i(\partial_0 +R_{ij}(\Theta) \eta_{ j}\partial_i) +  \frac{1}{2}m {\vec \eta^{2}} + \varepsilon(R(\Theta)_{ij}(i\partial_j +m\eta_j ))\right] \psi .
\end{eqnarray}
The kinetic piece of the angulon  Lagrangian consistent with time reversal and parity invariance is given by 
\beq
\label{KE}
L_{KE}= (\dot \Theta)^2+D^{ij} (\nabla_i \Theta)(\nabla_j \Theta),
\eeq
so the angulon is a ``type I" Goldstone, i.e. $E \sim p$.
Unlike the framid, the  angulon scaling  is not fixed by symmetry and its momentum scaling is determined
by the maximum momentum transfer consistent with the effective theory. i.e. the scattering of an electron
with an angulon should leave the electron near the Fermi surface to within $\lambda$ thus the angulon momentum 
scales as $\lambda$ and following the same arguments as above the field $\Theta(x) \sim \lambda^{1/2}$ in two spatial dimensions and as $\lambda$ in three.

Expanding the action (\ref{Spsi}) and keeping on the leading order piece we have

\begin{eqnarray}
S_{\psi}
&=&\int d^dx dt~\psi^\dagger\left[ i\partial_0+ i\vec \eta \cdot \vec \partial+ \vec v_F\cdot   (i\vec \partial +m \vec \eta)     +i\Theta v_F^i  \partial^j  \epsilon_{ij}\right] \psi .
\end{eqnarray}
We see that for $d=2$, the interaction with the angulon is relevant and thus destroys Fermi liquid behavior.
In $d=3$ it is classically marginal and the fate of Fermi liquid behavior is determined by the sign of the
beta function for this coupling.

Notice that the breaking of rotational symmetry does not effect the operator relation (\ref{Landau2}) imposed by the 
non-linearly realization of boost invariance. However, at least in two spatial dimensions, the Landau relation (\ref{Landau2}) is no longer justified, as the
angulon coupling becomes strong in the IR and quasi-particle picture breaks down.
In three dimensions it is possible that a perturbative result for the Landau relation could follow if the theory
remains weakly coupled.  In any case the operator constraint $(\ref{Landau2}$) must  hold for the system to
be boost invariant. However, in strong coupling it is not easy to deduce the physical ramifications. It would be interesting
to utilize this constraint to generate new prediction in systems with broken rotational symmetry. In particular it is
interesting to ask whether or not one can impose a DIHC to eliminate the angulon from the spectrum.

\subsection{The Stability of Goldstone Boson Mass Under Renormalization}

As can be seen from the actions (\ref{framonKE}) and (\ref{KE}), a Goldstone boson mass is forbidden despite the fact that the
Goldstone boson need not be derivatively coupled. There are no gapped Goldstones as a consequence of the
fact that there is no inverse Higgs mechanism for our chosen symmetry breaking pattern. If there is no anomaly then we should expect that
this masslessness should persist to all orders in perturbation theory, indeed it should hold non-perturbatively.
Vishwanath and Watanabe showed the cancellation of angulon mass correction at one loop \cite{VW} but they did not
consider the framon.
Given that we have constructed the full action, the all orders proof follows from the Ward identity.
Nonetheless is it  instructive to study the one loop case in order to distinguish the framon from the
angulon. The Goldstone mass can be read off by considering the quadratic piece of the effective
action generated by integrating out the electrons in a constant Goldstone background.
For the angulon we find
\beq
\Gamma[\theta] = i \frac{\int d\omega d^{d}p  Log\left[ \omega -\varepsilon (R(\theta) \vec p)\right]}{\int d\omega d^{d}p  Log\left[ \omega -\varepsilon (\vec p)\right]}
\eeq
which is independent of $\theta$ as a consequence rotational invariance of the measure. 
This result tells us that, at the level of the integrals, there must be an algebraic cancellation
between the two diagrams which contribute to the mass at one loop shown in figure 3.
Note this should NOT be expected for the framon, since boost symmetry breaking is sensitive
to the UV scale $E_F$, whereas the angulon only knows about the shape of the Fermi surface and
not its depth. Of course, boost invariance dictates the framon mass must vanish if we use
a boost invariant regulator, i.e. not a cut-off. The situation is analogous to the case of the dilaton whose mass corrections
vanish in dimensional regularization but necessitates counter-terms when using a cut-off. Such counter-terms
should not be considered fine tuning.

\section{Broken Conformal symmetry: Eliminating the non-Relativistic Dilaton}

As mentioned in the introduction, consequences of spontaneous breaking of conformal invariance in non-relativistic systems is unique as the non-relativistic kinetic term for the dilaton appears to be in tension with boost invariance \cite{oz}.
As such, we will study systems for which the broken symmetries are dilatations ($D$), special conformal transformations ($C$)  and boosts ($K_i$). The relevant commutators of the Schrodinger group (the non-relativistic conformal group) are
\bea
\label{HC}
[H,C]&=& i D
\eea
\bea
\label{IVH1}
  [P_i,C] &=&-iK_i\eea
as these relations imply a reduction in the naive number of Goldstones. Furthermore, note that (\ref{HC}) implies that if dilatations are broken then so are the special conformal transformations. The vacuum is parameterized via
\beq
U=e^{i P \cdot x} e^{-i \vec K \cdot \vec \eta}e^{-i C \lambda}e^{-i D \phi}.
\eeq
The algebra implies that both $\lambda$ and $\phi$ are redundant degrees of freedom. The ensuing vierbein is given by
\beq
\label{newvierbein}
E_0^0 =e^{-2 \phi}~~~E_0^i= \eta^i e^{-\phi}~~~ E_i^j= \delta_i^j e^{-\phi}.
\eeq
The gauge fields are 
\beq
\label{newgauge}
A_0= -\frac{1}{2}\vec \eta^2e^{2 \phi}~~~\vec A = -\vec \eta e^{\phi}
\eeq
and the covariant derivatives are given by
\bea
\nabla_j \eta^i &=& e^{2 \phi} (\lambda \delta_j^i+\partial_j \eta^i)   \\
\nabla_0 \eta^i&=&-e^{3\phi} (\dot \eta^i +\vec \eta \cdot \vec \partial \eta^i) \\
\nabla_0 \phi&=& -e^{2 \phi}(\lambda+\dot \phi  +\vec \eta \cdot \vec \partial \phi)\\ 
\nabla_i \phi &=& e^{\phi}\partial_i \phi  \\
\nabla_0 \lambda &=&  -e^{4\phi}(\dot \lambda+\vec \eta \cdot \vec \partial \lambda+\lambda^2) \\
\nabla_i \lambda &=& e^{3\phi}\partial_{i}\lambda
\eea

The invariance of these objects under boosts, dilatations and special conformal transformations follows by first determining the non-linear transformation
properties of the Goldstones via the relation
\beq
g U(\pi)= U(\pi^\prime,g) h(g,\pi),
\eeq
where $g \in G$ and $h \in H$.
Table 1 gives the resulting transformation properties of the Goldstones.

We see that there are two possible  inverse Higgs constraints coming  from setting the covariant derivatives in $(6.6)$ and $(6.8)$ to zero.
Linearizing yields the two possible inverse Higgs relations from $(6.6)$ and $(6.8)$
\bea
\label{invhiggs}
\lambda&=& - \dot \phi +...\nn \\
 \lambda&=& \frac{1}{3} \vec \partial \cdot \vec \eta+.....
\eea

Let us now address the question of the possible symmetry realizations.
We will see that no matter what path is chosen, the systems will not behave like a
canonical Fermi liquid \cite{us}.
 We may choose not to eliminate any Goldstones, however note that in this case, the $\lambda$ gets gapped as (6.10)
 is time reversal invariant and thus an allowed term in the action without squaring it.  This realization  includes two non-derivatively coupled Goldstones which would invalidate a
Fermi liquid description \cite{VW}.  
If we use one IHC then again we will have the same spectrum and the same conclusion is reached.
Finally we may consider using both constraints such that we equate
\beq
\dot \phi= \partial \cdot \eta
\eeq
which would lead to a theory which appears non-local.\footnote{This non-locality in EFT arises due to a poor choice of variables and is not in any sense fundamental since the underlying theory is local.} Thus, although we have two possible constraints we can only impose one
while maintaining locality. This is a consequence of the fact we have
\bea
[H,C]=iD ~~~~[C,P]=iK,
\eea
so that the two constraints are linked establishing the fact that the criteria for Goldstone elimination stated in (\ref{IH}) must be amended.
If two of the relation involve the same generator on the LHS then there is one fewer allowable constraint. We know of no other cases where this happens.
The final possibility is that we eliminate both $\eta$ and $\phi$  using DIHMs as discussed in the next section.

\begin{table}[h!]
  \centering
  \caption{Infinitesimal variation of Goldstones under broken charges}
  \label{tab:table1}
  \begin{tabular}{c|c|c|c}
  & $\vec K(\vec \beta)$ & $D(\alpha$) & $C(\rho$)\\
    \toprule
$\vec \eta$   & $-\vec \beta$ & $\alpha \vec \eta$ & $-\rho \vec x+t\rho\vec \eta$\\
$\phi$    &0 & $-\alpha$ & $-\rho t$\\
$\lambda$   & 0 & $2\alpha \lambda$ & $\rho-2 \rho t \lambda$\\
$\partial_{t}$ &-$\vec{\beta}\cdot\vec{\partial}$& $2\alpha\partial_{t}$ & $2t\rho\partial_{t}+\rho\vec{x}\cdot\vec{\partial}$\\
$\partial_{x}$ &0&$\alpha\partial_{x}$&$-t\rho\partial_{x}$\\
  \end{tabular}
\end{table}

\subsection{Consequence of Broken Conformal Symmetry via the DIHM}
To derive the relevant DIHCs we will again build the coset and treat both the dilaton and the framon as Lagrange multipliers. 
As in the previous cases, in two spatial dimensions this is not a choice as a  consequence of power counting and symmetry.
Notice that $\lambda$ will not play a role as it shows up neither in the vierbein nor the connection.
We have already written down the
most general boost invariant interaction in (\ref{action1}) which we now amend using the new version of
the vierbein and gauge field (\ref{newvierbein}, \ref{newgauge}). The invariant action for the quasi-particle is given by
\begin{eqnarray}
\label{action2}
S_0
&=&\int \frac{d^{d}pdt}{(2\pi)^{d}} e^{2\phi}~{\psi}_{{ \vec {\bf p}} }^\dagger(t)\left[ (ie^{-2\phi}\partial_0-e^{-2\phi}\vec \eta\cdot { \vec {\bf p}}  + \tilde{\varepsilon}(e^{-\phi}(\vec p+m\vec \eta))+\mu_{\text{F}}\right] {\psi}_{{ \vec {\bf p}} }(t) ,\nn \\
\end{eqnarray}
Here the energy functional $\tilde{\varepsilon}(p)$ is the energy of the quasi-particle measured from the Fermi surface since we have explicitly included the chemical potential $\mu_{F}$ in the action. For notational convience we will drop the explicit factor of $\mu_{F}$ and redefine the energy functional as $\varepsilon(p)=\mu_{F}+\tilde{\varepsilon}(p)$. As far as the interactions are concerned we have
\begin{eqnarray}
\label{qpint2}
S_{int}&=&\frac{1}{2} \int\prod_{a=1}^4\frac{d^{d}p_{a}}{(2\pi)^{d}}dt \delta^{(d)}(p_{1}+p_{2}-p_{3}-p_{4}) e^{(2-d)\phi} ~g({{e^{-\phi} \vec{\bf p}_i}}+e^{-\phi}m \vec{\eta}_i,e^{-\phi}\mu){\psi}^\dagger_{{\vec{\bf p}_1}}(t) {\psi}_{{\vec{\bf p}}_2}(t) 
{\psi}^\dagger_{\vec{\bf p}_3}(t){\psi}_{\vec{\bf p}_4}(t). \nn \\
\end{eqnarray}
Here we have also introduced the renormalization scale $\mu$ in the coupling. The Landau relation (Eq.(\ref{GL})) which ensures boost invariance remains unchanged but we generate a new constraint by
 setting $\eta$ to zero and varying the action (Eq.(\ref{action2}) and Eq.(\ref{qpint2}))
with respect to  $\phi$.
Expanding Eq.~(\ref{action2}) and Eq.~(\ref{qpint2}) to leading order in $\phi$, 
\begin{eqnarray}
\label{qpint3}
S^{\phi}&=&\sum_{\vec{\bf k}}\int d^dp dt~~ \phi~\psi^{\dagger}_{\vec{\bf p}}(t)\left[2 \varepsilon(p)- p^i  \frac{\partial \varepsilon}{\partial p_i} \right]\psi_{\vec{\bf p}}(t) \nn \\
&+&\frac{1}{2}\prod_{a=1}^{4}\int d^dp_{a} dt~\phi\left[(2-d)g(\vec{\bf p}_{i},\mu)-\vec{\bf p}_{i}\cdot\frac{\partial{g(\vec{\bf p}_{i},\mu)}}{\partial\vec{\bf p}_{i}}-\mu\frac{\partial{g(\vec{\bf p}_{i},\mu)}}{\partial\mu}\right]\psi^{\dagger}_{\vec{\bf p}_{1}}(t)\psi_{\vec{\bf p}_{2}}(t)\psi^{\dagger}_{\vec{\bf p}_{3}}(t)\psi_{\vec{\bf p}_{4}}(t) \nn \\
\end{eqnarray}
The constraint follows from imposing $\frac{\delta{S^{\phi}}}{\delta\phi}=\mathcal{O}_{\phi}=0$.
\bea
\label{dilaton}
\mathcal{O}_{\phi}&=&\sum_{\vec{\bf k}}\int d^dp dt~~\psi^{\dagger}_{\vec{\bf p}}(t)\left[2 \varepsilon(p)- p^i  \frac{\partial \varepsilon}{\partial p_i} \right]\psi_{\vec{\bf p}}(t) \nn \\
&+&\frac{1}{2}\prod_{a=1}^{4}\int d^dp_{a} dt~\left[(2-d)g(\vec{\bf p}_{i},\mu)-\vec{\bf p}_{i}\cdot\frac{\partial{g(\vec{\bf p}_{i},\mu)}}{\partial\vec{\bf p}_{i}}-\mu\frac{\partial{g(\vec{\bf p}_{i},\mu)}}{\partial\mu}\right]\psi^{\dagger}_{\vec{\bf p}_{1}}(t)\psi_{\vec{\bf p}_{2}}(t)\psi^{\dagger}_{\vec{\bf p}_{3}}(t)\psi_{\vec{\bf p}_{4}}(t)=0 \nn \\
\eea

Let us now see if a Fermi liquid description is consistent with these constraints.
Given our assumption of rotational invariance and the notion of a  well defined Fermi surface, the marginal coupling  is only a function 
of the angles which are scale invariant. Thus the second term in the last line of (\ref{dilaton}) vanishes, and,  as such, if we take
the one particle matrix element we see that the quadratic and quartic terms must vanish
separately since the quadratic term will depend upon the amplitude of the incoming external 
momentum and the quartic will not. In three dimensions we see that the coupling has power law running which 
is inconsistent with Fermi liquid theory, and in two dimensions the theory is free. Thus we conclude that: {\it fermions at unitarity are not properly described by Fermi liquid theory}. 

We can also consider how these symmetry constraints can be utilized if we assume that the microscopic
theory is defined via the action (\ref{qpint}) (i.e. its is not an effective theory) 
as done in simulations. 
In this case since there is no restriction to forward scattering  there is no mechanism by which
the  quadratic term can cancel with the quadratic for all choices of states.
Then taking the one particle matrix element of
(\ref{dilaton})   we have the constraints
\beq \boxed{
\varepsilon = \frac{p^2}{2m^\star} } \eeq
and
\beq
\label{beta2}
\boxed{
0=(2-d)g(\vec{\bf p}_{i},\mu)-\vec{\bf p}_{i}\cdot\frac{\partial{g(\vec{\bf p}_{i},\mu)}}{\partial{\vec{\bf p}_{i}}} -\beta(g).}
\end{equation}
For S-wave scattering ($g(p,\mu)$=$g(\mu)$), $m=m^\star$ due to the Landau condition in Eq.(\ref{Landau1})
and $(2-d)g(\mu) =\beta(g)$.
For higher angular momentum channels, Eq.(\ref{beta2}) gives us the beta function to all orders.

\section{Conclusions}
The predicitve power of symmetries is not lost when the ground state is not invariant. The symmetries are simply realized in a non-linear fashion. For internal symmetries Goldstone boson appears which saturate the relevant Ward identities.
 When space-time symmetries are broken, new pathways to symmetry realization arise. We have
shown in particular, that the inverse Higgs mechanism which leads to a reduction in the number
of Goldstones  can be generalized. We introduced the
notion of a dynamical inverse Higgs mechanism (DIHM) whereby a strong operator constraint (DIHC)  is imposed
which enforces symmetry realization.
A simple example of a DIHC was presented here in the context
of Fermi liquid theory. The missing boost Goldstone is seen to be absent as it is unnecessary  once the DIHC is imposed.
In this case the constraint leads to the Landau relation which relates the coupling to the effective mass.
This in itself does not explain why we know of no systems which manifest a boost Goldstone, as
in three dimensions the framon (the boost Goldstone) path to symmetry realization seems to be perfectly consistent.
Thus there is no a priori reason why we would expect no such systems to arise in nature. However, as we have shown, 
the framon is non-derivatively coupled, thus it is natural to expect that in the IR, strong dynamics will
set in and change the relevant degrees of freedom. As such, the framon could be hiding under the shroud of
strong coupling.

In general we do not know a priori if a given DIHC can be satisfied. In the case of the breaking of boost invariance
we showed that the constraints force all low energy interactions to be relegated to particular kinematic
configurations, i.e. back-to-back (BCS) and forward scattering. At the same time, canonical Fermi liquid theory
tells us that these are the only possible marginal interactions based on power counting. Clearly this is no coincidence.

We then presented an example of a DIHC which can not be satisfied by studying the realization of the
Schrodinger group broken by a Fermi sea. We showed how the broken symmetries can be realized by
the inclusion of a dilaton and a framon.
Using the boost DIHC (Landau relation) to eliminate the associated
Goldstone leads to another DIHC which allows for the elimination of the dilaton as well. In two dimensions
the DIHC can be satisfied and is the only path to symmetry realization while in three dimension 
the DIHC leads to a constraint that contradicts the Lagrangian dynamics. The full Schrodinger symmetry can be 
 consistently realized by including a dilaton and a framon, while the Goldstone of special conformal transformation
gets gapped. However, the ensuing theory is not consistent with a Fermi liquid description as the non-derivative coupling
of the Goldstone would leads to (at least) marginal Fermi liquid behavior\cite{us}.

\appendix
\section{Landau Relation from Galilean algebra}
The Landau relation can also be derived (similar to Landau's original derivation) by demanding that the Fermi Liquid action, without including the boost Goldstone, should be Galilean boost invariant. This is equivalent to satisfying the Galilean algebra by using the Noether charges constructed from the Fermi Liquid action. The only commutator of the Galilean algebra we need to satisfy is $[H,G_{i}]=iP_{i}$  where $G_{i}$ is the generator of Galilean boost, H is the Hamiltonian and $P_{i}$ is the momentum operator. In terms of the quasi-particle fields, these operators are given by
\begin{align}
H=&\int d^{d}p~\psi^{\dagger}_{p} \varepsilon(p) \psi_{p}+\prod_{i}\int d^dk_i~\frac{g(k_i)}{2}\psi^\dagger_{k_1} \psi^\dagger_{k_2} \psi_{k_3} \psi_{k_4} \delta^{(d)}(k_1+k_2-k_3-k_4)\nonumber \\ 
G_{i}=&t\int d^{d}p~\psi^{\dagger}_{p} p_{i} \psi_{p}-im \int d^{3}p~\psi^{\dagger}_{p} \partial_{i} \psi_{p} \nonumber \\
P_{i}=&\int d^{d}p~\psi^{\dagger}_{p} p_{i} \psi_{p}
\end{align}

Using anti-commutation relation $\{\psi_{p},\psi^{\dagger}_{p'}\}=\delta^{d}(p-p')$ and satisfying $[G_{i},H]=iP_{i}$, we get back the operator relation in (\ref{GL}).
\section{Landau Relation from Poincar\'e algebra}
Here we derive the Landau relation for a relativistic Fermi liquid from current algebra.
The same result can be reached by using the relativistic coset construction.
The derivation of Landau relation for the relativistic Fermi liquids \cite{Baym} is a little more involved then compared to the Galilean case. The commutator we need to satisfy is still $[H,K_{i}]=iP_{i}$ where $K_{i}$ is the generator of the Lorentz boost's but the Noether charges are different from their Galilean counterparts. Denoting $H_{0}$ as the free Hamiltonian and V as the interaction
\begin{align}
H_{0}=&\int d^{d}p~\psi^{\dagger}_{p}\varepsilon(p) \psi_{p} \nonumber \\
V=&\prod_{i}\int d^dk_i~\frac{g(k_i)}{2}\psi^\dagger_{k_1} \psi^\dagger_{k_2} \psi_{k_3} \psi_{k_4} \delta^{(d)}(k_1+k_2-k_3-k_4)\nonumber \\ 
k_{i}=&t\int d^{d}p~\psi^{\dagger}_{p} p_{i} \psi_{p}- i\int d^{d}p \partial_{i}\psi^{\dagger}_{p}\varepsilon(p)\psi_{p}\nonumber \\
W_{i}=&\prod_{i}\int d^dk_i~g(k_i)\partial_{i}\psi^\dagger_{k_1} \psi^\dagger_{k_2} \psi_{k_3} \psi_{k_4} \delta^{(d)}(k_1+k_2-k_3-k_4)\nonumber \\  
P_{i}=&\int d^{d}p~\psi^{\dagger}_{p} p_{i} \psi_{p}
\end{align} 
where $W_{i}$ is the correction to the boost operator due to presence of interactions. This is because $K_{i}=tP_{i}-\int x_{i}T^{00}$ where $T^{\mu\nu}$ is the energy momentum tensor, and $T^{00}$ is the sum of free and the interacting Hamiltonian density. So we defined $K_{i}=k_{i}+W_{i}$ where $k_{i}$ is from free part of the Hamiltonian density and $W_{i}$ is due to interactions. The Poincar\`e algebra condition now becomes
\begin{equation}
[H_{0}, k_{i}]+[H_{0}, W_{i}]+[V, k_{i}]+[V, W_{i}] = iP_{i}
\end{equation} 
Assuming weak interactions between quasi-particles and neglecting terms of $\mathcal{O}(g^2)$, (taking one particle matrix elements for a state with external momentum, k)
\begin{equation}
\varepsilon(k)\frac{\partial}{\partial{k_{i}}}\left(\varepsilon(k)+\varepsilon(k)~\int d^{d}p \langle0|\psi^{\dagger}_{p}\psi_{p}|0\rangle g(p,k)\right)+\int d^{d}p \langle0|\psi^{\dagger}_{p}\psi_{p}|0\rangle \frac{\partial}{\partial{p_{i}}}(\varepsilon(p)g(p,k))=k_{i}
\end{equation}
For forward scattering $g(p,k)=g(\cos \theta)$ where $\theta$ is angle between $\vec{p}$ and $\vec{k}$ and $\langle0|\psi^{\dagger}_{p}\psi_{p}|0\rangle=\Theta(p_{F}-p)$ to leading order in g. Using $\varepsilon(k_{F})=\mu$, where $\mu$ is the chemical potential and the definitions of effective mass, $m^{*}$ and the density of states at the Fermi surface, $D(\mu)$,
\begin{equation}
m^{*}=\mu\left(1+\frac{1}{3}G_{1}\right)
\end{equation}    
where we assumed $d=2$ and expanded the coupling function $g(\theta)$ in Legendre polynomials , $g(\theta)=\sum_{l}g_{l}P_{l}(\cos \theta)$ and defined $G_{l}=D(\mu)g_{l}$.
\section*{Acknowledgements}
This work supported by the DOE contracts DOE DE-FG02-04ER41338 and FG02- 06ER41449. The authors thank Riccardo Penco for comments on the manuscript.

\end{document}